\documentclass[aps,amssymb,twocolumn,superscriptaddress,nofootinbib]{revtex4}
\usepackage{setspace}
\usepackage{graphicx}
\usepackage{psfrag}

\newcommand{\pstar}{{p^*}}
\newcommand{\bstar}{{b^*}}
\newcommand{\bbar}{{\bar{b}}}
\newcommand{\Ebar}{{\bar{E_r}}}
\newcommand{\kstar}{{k^*}}
\newcommand{\Screll}{{\mathcal L}}

\def\be{\begin{equation}}
\def\ee{\end{equation}  }
\def\bea{\begin{eqnarray}}
\def\eea{\end{eqnarray}  }

\begin{document}
\title{Black Hole Mergers and Unstable Circular Orbits}

\author{Frans Pretorius}
\affiliation{Department of Physics, Princeton University, Princeton, NJ 08540}
\author{Deepak Khurana}
\affiliation{Indian Institute of Technology, Kharagpur}
\begin{abstract}
We describe recent numerical simulations of the merger 
of a class of equal mass, non-spinning, eccentric
binary black hole systems in general relativity. We show that
with appropriate fine-tuning of the initial conditions to
a region of parameter space we denote the {\em threshold
of immediate merger}, the binary enters a phase of close interaction
in a near-circular orbit, stays there for an amount of time proportional
to logarithmic distance from the threshold in parameter
space, then either separates or merges to form a single
Kerr black hole. To gain a better understanding of this
phenomena we study an analogous problem in the evolution
of equatorial geodesics about a central Kerr black hole. A similar
{\em threshold of capture} exists for appropriate classes of
initial conditions, and tuning to threshold
the geodesics approach one of the unstable circular geodesics
of the Kerr spacetime. Remarkably, with a natural
mapping of the parameters of the geodesic to that of the
equal mass system, the scaling exponent describing the
whirl phase of each system turns out to be quite similar.
Armed with this lone piece of evidence that an approximate correspondence
might exist between near-threshold evolution of geodesics and
generic binary mergers, we illustrate how this information can
be used to estimate the cross section and energy emitted
in the ultra relativistic black hole scattering problem. This
could eventually be of use in providing estimates for the
related problem of parton collisions at the Large Hadron Collider
in extra dimension scenarios where black holes are produced.
\end{abstract}

\maketitle

\section{Introduction}

Gravitational physics is entering an exciting era. The construction
of gravitational wave detectors that are expected to be sensitive
enough to observe many astrophysical phenomena where strong-field
gravity plays an important role should teach us much about
the cosmos and the structure of spacetime. Suggestions
that we may live in a universe with more than 3 spatial
dimensions~\cite{arkani_hamed_et_al,randall_sundrum} offer 
the intriguing possibility that
the Planck scale could be within reach of energies
attainable by the Large Hadron 
Collider (LHC)~\cite{banks_fischler,giddings_thomas,dimopoulos_landsberg}. 
This implies
that the LHC may be able to probe the quantum gravity regime,
and that black holes could be produced in substantial quantities
by the particle collisions. Similarly, cosmic ray collisions
with the earth would produce black holes~\cite{feng_shapere}, and this
may be detected with current or near-future cosmic ray 
experiments~\cite{landsberg}. Understanding
the nature of black hole collisions within the theory
of general relativity will be important in describing
and interpreting many of these fascinating phenomena,
if detected. 

The past couple of years has witnessed several
breakthroughs in numerical relativity\cite{paper2,utb,nasa}, allowing for the
solution of the field equations describing black hole mergers 
in many situations. However a vast
region of interesting parameter space is unexplored, and it will
be several years at least before a decent understanding of
black hole collisions is achieved.
In this paper we describe simulations of the merger of
a class of equal mass black hole binaries on initially eccentric orbits.
The orbits can be labeled with a one parameter family
$k$ loosely related to the initial tangential velocities
of each black hole. We find intriguing behavior tuning
$k$ to what we call the {\em threshold of immediate merger} separating
evolutions where the black holes either merge or do not during
their first close encounter. The resultant evolution
becomes exponentially sensitive to the initial parameter,
and the binaries exhibit a period of ``whirl'' type 
behavior similar to that seen in geodesic 
motion~\cite{zoom_whirl,zoom_whirl_b}, orbiting rapidly in a near-circular configuration.
Remarkably, significant amounts of gravitational
radiation ($\approx1.0-1.5\%$ of the rest mass energy {\em per orbit}) are still 
being emitted in this regime. Furthermore, based on the coordinate
separation\footnote{Though this is not a gauge invariant quantity,
though comparisons between
the extracted waveform and that estimated using the quadrupole
formula suggests the coordinate motion of the apparent horizons
is quite well adapted to describing the situation\cite{buonanno_et_al}.}, 
the binaries are orbiting within what would be the
innermost stable circular geodesic of a Kerr spacetime with
angular momentum equal to that of the black hole that forms
in the merger case.

We show that the above behavior for equal mass binaries
is analogous to evolution
of similar classes of geodesics in a black hole background
spacetime. Namely, if we define a one parameter family
of equatorial geodesics and tune to the threshold of capture, at threshold
the geodesic will approach one of the unstable circular
geodesics of the background spacetime, regardless of whether the
initial orbit is classified as elliptic or hyberbolic.
Furthermore, if we calculate the instability (or Lyapunov) exponent
of the orbits near the unstable circular orbits, we find numbers
that are similar to that observed in this fully non-linear
equal mass case.

The binary black hole merger simulations described here are in the rest-mass
dominated regime of the problem. The close analogy with
geodesic motion allows us to {\em speculate} about what might happen
in the kinetic energy dominated regime. This is relevant to the
ultra-relativistic black hole scattering problem and thus might have
application to the LHC if the Planck scale is below the maximum
energies probed by the parton collisions. In particular, by 
finding the critical impact parameter and stability exponent
of geodesic motion, estimating the energy and angular momentum
loss while the geodesic is in the whirl phase, and providing
an estimate of the energy emitted for the head-on collision
case in the full problem, one can come up with an estimate
of the energy emitted to gravitational waves as a function
of impact parameter. We also speculate that at threshold,
{\em all} of the kinetic energy of the system is converted 
to gravitational waves, which can be an arbitrarily large
fraction of the total energy.

The outline of the rest of the paper is as follows.
In Sec.\ref{sec_method} we summarize the numerical code; in Sec.\ref{sec_threshold}
we describe the simulation results; in Sec.\ref{sec_geod} we describe
the geodesic analog; in Sec.\ref{sec_disc} we discuss all the results,
speculating about what may happen beyond the equal mass, non-spinning case,
and how the results might carry over to the kinetic energy dominated
regime and be applied to the BH scattering problem; in Sec.\ref{sec_conclude}
we provide some concluding remarks and a discussion of future related 
work. The technique we use for geodesic integration is described
in the appendix.

\section{Overview of the equations and solution method}\label{sec_method}

The Einstein Field Equations (EFE) in generalized harmonic form have
been discussed in much detail 
elsewhere~\cite{friedrich,friedrich2,garfinkle,szilagyi_et_al,
szilagyi_winicour,
paper1,cqg_review,new_lindblom_et_al,babiuc_et_al,babiuc_et_al_2,
szilagyi_et_al_b,rinne,alcubierre_et_al,rinne_stewart,gourgoulhon} 
so for brevity here we simply list the equations and briefly
mention the numerical code solving these equations.

\subsection{formalism}

We solve for a spacetime described by the line element $ds$ with metric tensor $g_{ab}$
and coordinates $x^a=(t,x,y,z)$\footnote{We use a comma ($,$) to denote partial 
differentiation, and we use units where Newton's constant $G=1$ and 
the speed of like $c=1$.}
\be
ds^2=g_{ab} dx^a dx^b,
\ee
using the EFE in generalized harmonic (GH) form with {\em constraint damping}
\cite{gundlach_et_al,lambda_ref}:
\bea
\frac{1}{2} g^{cd}g_{ab,cd} 
+ g^{cd}{}_{(,a} g_{b)d,c}
+ H_{(a,b)} \nonumber
- H_d \Gamma^d_{ab} \\
+ \Gamma^c_{bd}\Gamma^d_{ac} 
+ \kappa\left(n_{(a} C_{b)} - \frac{1}{2}g_{ab} n^d C_d\right) \nonumber\\
= - 8\pi\left(T_{\alpha\beta}-\frac{1}{2}g_{\alpha\beta} T\right)\label{efe_h_cd}.
\eea
In the above, $\Gamma^c_{ab}$ are the Christoffel symbols
\be\label{christoff}
\Gamma^{c}_{ab}=\frac{1}{2}
g^{ce}\left[g_{ae,b}+g_{be,a}-g_{ab,e},
\right]
\ee
$T_{ab}$ is the stress energy tensor with trace $T$,
$H_a$ are the GH {\em source functions} defined via
\be\label{gharm_def}
\Box x^c = H^c,
\ee
$\kappa$ is a parameter multiplying the constraint damping terms,
$n_a$ a timelike vector, and $C_a$ are the constraints:
\be
C^a \equiv H^a - \Box x^a. \label{c_def}
\ee
Any solution to the Einstein equations must have $C^a=0$; in a consistent numerical
evolution the constraints will be zero to within the truncation
error of the numerical scheme. For $n^a$ we use the usual
hypersurface unit normal vector
$n^a=1/\alpha(\partial/\partial t)^a$, where $\alpha$ is called the lapse function.

We couple in a massless scalar field $\Phi$ as the matter source,
which satisfies the wave equation
\begin{equation}\label{phi_eom}
\Box \Phi = 0,
\end{equation}
and has a stress energy tensor
\begin{equation}\label{set}
T_{ab} = 2 \Phi_{,a}\Phi_{,b} - g_{ab} \Phi_{,c}\Phi^{,c},
\end{equation}

The following equations are used to evolve the source functions:
\be
\Box H_t = - \xi_1 \frac{\alpha-1}{\alpha^\eta} + \xi_2 H_{t,\nu} n^\nu\label{t_gauge}, \ \ \ H_i=0.
\ee
This equation 
for $H_t$ is a damped wave equation with a forcing function
designed to prevent the lapse $\alpha$ 
from deviating too far from its Minkowski value of 
$1$. The parameter $\xi_2$ controls the damping term,
and $\xi_1,\eta$ regulate the forcing term. The particular
parameter values are as described in \cite{cqg_review}.

\subsection{Boosted Scalar Field Initial Data}

For initial conditions we use two boosted scalar field
pulses, as described in detail in \cite{cqg_review}. These pulses
very quickly undergo gravitational collapse and form a pair
of black holes in a bound orbit,
with most ($\approx 85\%$) of the scalar field energy falling into
the black holes, the rest radiating away on roughly the light-crossing
time scale of the orbit. At the initial time we assume the corresponding
spatial geometry is conformally flat and maximal, and solve the Hamiltonian
and momentum constraint equations together with the maximal slicing condition 
for self-consistent initial conditions. For the remaining coordinate
degrees of freedom we choose the spacetime slice to be harmonic
at $t=0$.

\subsection{Numerical code}
The evolution code, described in detail in \cite{paper1}, uses second
order accurate finite difference discretization with adaptive
mesh refinement, a coordinate system compactified to spatial infinity
and excision techniques to deal with the singularities inside of the
black holes. Properties of the black holes are measured using 
apparent horizon (AH) properties, and gravitational wave information
is extracted using the Newman Penrose formalism with the extraction
radius placed a distance of $50m$
from the origin, where $m=m_1+m_2$ is the total, initial AH mass
of the individual black holes measured after the majority of scalar
field energy has been accreted.

\section{The threshold of ``immediate'' merger}\label{sec_threshold}

Here we present one of the main results of this paper,
namely we give evidence that there are regions in the parameter space
of the two body problem in full general relativity where the black holes
evolve toward an {\em unstable circular orbit}, remain in 
that configuration for an amount of time sensitively
related to the initial conditions, then either plunge
toward coalescence or separate. In the case where the
black holes separate after the circular motion they could
possibly merge at some time in the future.
These regions in parameter space can be found by examining
suitable one parameter families $p$ of initial conditions,
where $p_i < p < p_s$ smoothly interpolates between  
an evolution with $p=p_i$ where a merger occurs promptly within $t=t_i$,
and an evolution with $p=p_s$ where after some amount of time $t_s \gg t_i$
no merger has occurred. The unstable circular orbit is
approached near {\em the threshold of immediate merger}
at $p\approx\pstar$, where for $p<\pstar$ merger occurs
promptly, while for $p>\pstar$ it does not\footnote{Note that depending
on the parameter that is being varied the magnitude
of $p$ relative to $\pstar$ denoting prompt merger can
be inverted.}. The number of
orbits $n$ observed near threshold is found to scale as
\begin{equation}\label{n_gamma}
e^{n} \propto |p-\pstar|^{-\gamma}.
\end{equation}
Note that due to the energy loss via gravitational radiation
the threshold cannot be ``sharp'', i.e. if the time $t_m(p)$ to
merger is plotted as a function of $p$, this will {\em not} have a
discontinuous step at $p=\pstar$. There will be a maximum
number of orbits $N$ for a given class of initial conditions,
and what from a distance might appear like a step function will 
be resolved into a smooth transition over a region of
size $\delta p \approx e^{-N/\gamma}$.

We cannot {\em prove} some of the statements made in the
preceding paragraph, in particular since the full numerical
simulations are so computationally expensive we have
only studied one class of initial conditions in detail.
However, these simulation results, the striking
similarity between them and the geodesic analogue
presented in the next section, and trying to understand
what must happen at a generic threshold as discussed in Sec.\ref{sec_disc}
provides quite a compelling case for this behavior.

\subsection{Simulation Results}

Here we describe results from the evolution of a
class of scalar field collapse
binary (SFCB) simulations discussed before in \cite{cqg_review}. 
The new simulation data presented here includes several
higher resolution runs tuned closer to the threshold
of immediate merger, and so now we have significantly more confidence that 
we are converging to this phenomenon in the two body problem.
Below we give a brief summary of the initial conditions, and
focus on evolution results of relevance to classifying and
understanding the immediate-merger-threshold scenario. More 
background details can be found in \cite{cqg_review}.

For initial conditions we begin with two 
identical boosted scalar field distributions,
one located at a coordinate location of $(x,y,z)=(4.45m,0,0)$ and 
given a boost with boost parameter $k$ in the positive-$y$ 
direction, while the other is located at
$(x,y,z)=(-4.45m,0,0)$ and given a boost $k$ in the negative-$y$
direction (the proper separation is initially $10.8m$). 
The scale $m$ here is the sum of apparent horizon masses
measured around $t=15m$ after evolution has begun, which is
after essentially all of the collapsing scalar field energy has
either accreted into the newly formed black holes or is on its
way to escaping to infinity. Approximately $85\%$ of the initial
scalar field energy falls into the black holes.
Thus $k$ labels our family of initial conditions. Note that
$k$ is related to though not exactly the same as the
initial velocities the black holes will have.
With $k=0$ we have a head-on collision, i.e. a prompt 
merger, while for $k$ sufficiently large the black holes
are deflected but fly apart. Thus $k$ describes an appropriate
family to study the threshold of immediate merger.

\begin{figure}
\begin{center}
\includegraphics[width=8.5cm,clip=true]{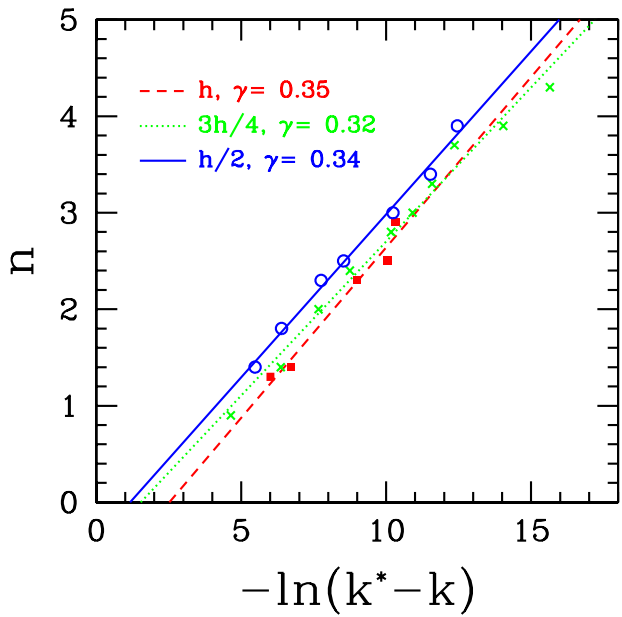}
\end{center}
\caption{The number of orbits $n$ versus logarithmic
distance of the initial boost parameter $k$ from the
immediate merger threshold $\kstar$, for evolutions
that did result in a merger. Results from the three
resolutions are plotted, and a least-squares
fit to each set of data assuming the relation (\ref{n_gamma}).
}
\label{fig_gamma_merge}
\end{figure}

\begin{figure}
\begin{center}
\includegraphics[width=8.5cm,clip=true]{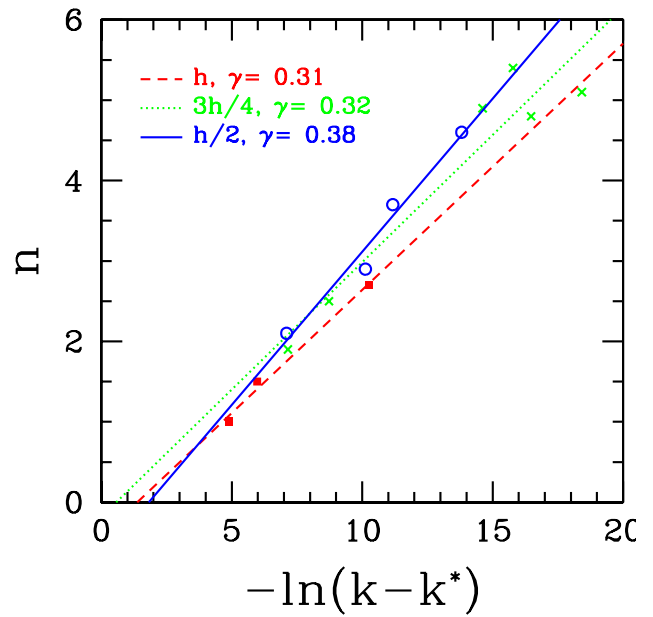}
\end{center}
\caption{ Data as in Fig. \ref{fig_gamma_merge}, though
here from evolutions that did {\em not} merge during
the time of the simulation (i.e. $k>\kstar$).
}
\label{fig_gamma_disp}
\end{figure}

\begin{figure}
\begin{center}
\includegraphics[width=8.5cm,clip=true]{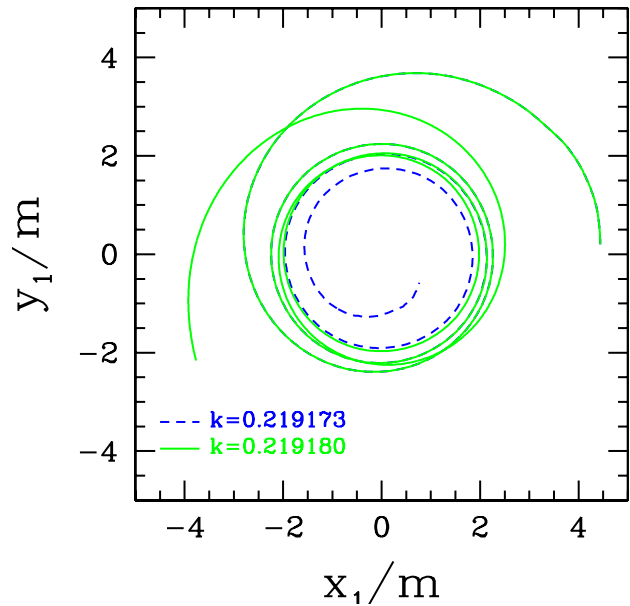}
\end{center}
\caption{Plots of the orbital motion from the two
higher resolution simulations ($h/2$) tuned closest
to threshold (only the coordinate motion of a single
black hole---initially at positive $x$---is shown for clarity). 
The dashed curve is
the case resulting in a merger, and the curve ends
once a common apparent horizon is first detected, while
for the solid curve the black holes separate again and
here the curve ends when the simulation was stopped.
}
\label{orbit_example}
\end{figure}

\begin{figure}
\begin{center}
\includegraphics[width=8.5cm,clip=true]{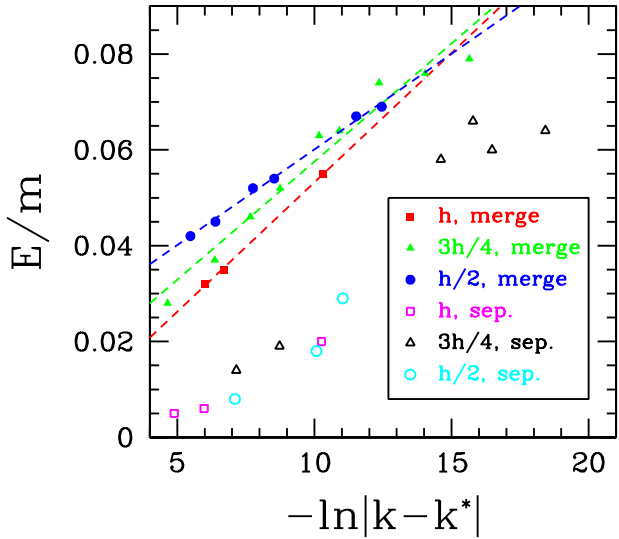}
\end{center}
\caption{
The total energy radiated in gravitational waves
plotted as a function of logarithmic distance
from the immediate merger threshold. We have
overlayed the data from both super and sub critical cases,
though for clarity have only added the linear
regression line for the cases that merged.
}
\label{fig_k_E}
\end{figure}

\begin{figure}
\begin{center}
\includegraphics[width=8.5cm,clip=true]{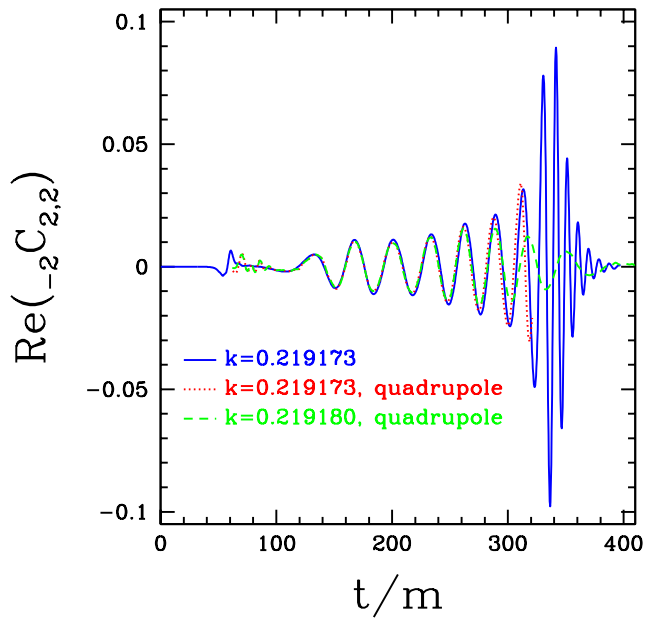}
\end{center}
\caption{The gravitational waves emitted during
a merger event. Here we show the real part
of the dominant spin weight -2, $\ell=2$, $m=2$ 
spherical harmonic component of $\Psi_4$. The solid
curve is the merger case tuned closest to threshold
(from the higher resolution simulations).
The dotted curve was computed
by taking the coordinate motion of the AH's from
this simulation (see Fig.\ref{orbit_example}), assuming they 
represent the locations
of point particles of mass $m/2$, and plugging the 
result into the quadrupole formula; this waveform
ends once a common horizon is formed. The dashed curve
shows the same information as the dotted curve, but
here the data is from the non-merger case tuned
closest to threshold. Note that the two curves from the
quadrupole formula were shifted in time to account
for the propagation time to the sphere at $r=50m$ where
$\Psi_4$ was measured.
}
\label{fig_Lm2_c22r}
\end{figure}

\begin{figure}
\begin{center}
\includegraphics[width=8.5cm,clip=true]{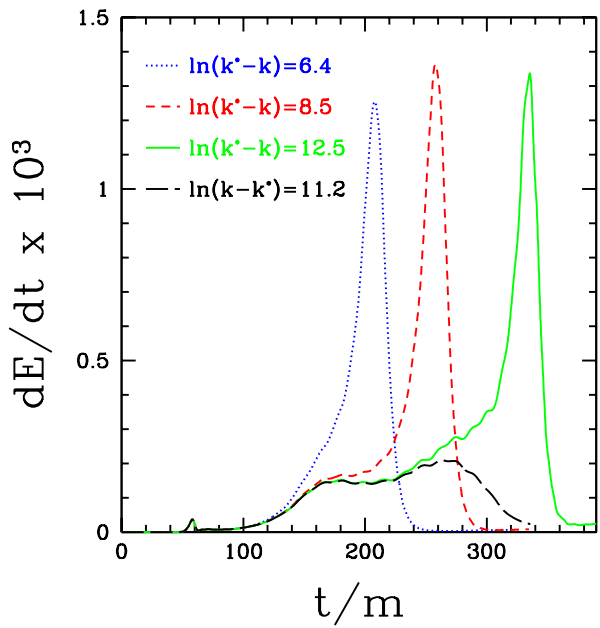}
\end{center}
\caption{The energy flux in gravitational waves from 
four ($h/2$ resolution)
simulations; three resulted in mergers, the fourth (long-dashed
curve) not.
}
\label{fig_dE_dt}
\end{figure}

Fig.~\ref{fig_gamma_merge} shows $n(k)$ (\ref{n_gamma})
for the cases that merged ($k<\kstar$), while Fig.~\ref{fig_gamma_disp}
shows $n(k)$ for the cases that separated after the
initial whirling motion. In the former plot $n$
is calculated as the total phase angle $\phi$ divided
by $2\pi$ that one of the black holes moves through
before a common AH is detected, while in the later case
is the total $\phi$ evolution divided by $2\pi$ undergone by
the black hole in returning to its original coordinate distance
from the origin. An example of the orbital trajectories
is shown in Fig. \ref{orbit_example}.
Fig. \ref{fig_k_E} shows the total energy radiated in gravitational
waves from these  simulations, Fig. \ref{fig_Lm2_c22r} shows
a sample of the waveform from a couple of the simulations measured
using both the Newman-Penrose scalar $\Psi_4$ and the 
quadrupole formula, and Fig. \ref{fig_dE_dt} shows the energy
flux (calculated using $\Psi_4$) from four sample simulations.

From the data shown in Fig.\ref{fig_gamma_merge} we estimate
$\gamma=0.35\pm0.03$ for this family of initial data.
It is difficult to calculate the uncertainty in this
quantity. In theory convergence testing should
be sufficient, though here each resolution has
a different number of points, so the intrinsic
error in a linear regression fit will be different.
Furthermore, we are assuming (\ref{n_gamma}) holds,
and it most certainly does not exactly. Also, since
each set of simulations span a different range
in $k-\kstar$ this will cause some variation
in the measured $\gamma$ in addition to truncation
and small sample size errors. We have therefore
simply taken the uncertainty to be the difference
in $\gamma$ between the medium and higher resolution
simulations (which is roughly what it would be
if only truncation errors were responsible for
the differences). A summary of the three characteristic
resolutions used is listed in Table \ref{tab_res}.

\begin{table}
\begin{tabular}[t]{| l l || c | c | c |}
\hline
 & ``Resolution'' & wave-zone res.& orbital-zone res.& BH res.\\
\hline
\hline
 & h     & $1.7  m$ & $0.23 m$  & $ 0.057 m$ \\
 & 3/4 h & $1.3  m$ & $0.17 m$  & $ 0.043 m$ \\
 & 1/2 h & $0.85 m$ & $0.12 m$  & $ 0.029 m$ \\
\hline
\end{tabular}
\caption{The three sets of characteristic resolutions used in
simulation results presented here, where each resolution 
is labeled relative to the coarsest resolution $h$.
The grid is adaptive with a total of 8 levels of refinement,
and the coordinate system is compactified. The wave zone
is defined to be at $r=50 m$, the orbital zone within about $r=10 m$
and the black hole zone is within $2-3m$ of each apparent horizon,
where $m$ is the initial sum of apparent horizon masses, measured
at $t=15m$ to account for early scalar field accretion.
At this time the coordinate radius of each apparent horizon
is $\approx 0.52m$; thus for the highest (lowest) resolution 
case roughly 36 (18) grid points span each horizon.
A CFL (Courant-Friedrichs-Lewy) factor
of $0.15$ was used in all cases. 
}
\label{tab_res}
\end{table}

\section{The geodesic analogue}\label{sec_geod}

To gain insight into what is happening 
when tuning to the threshold of immediate merger
it is useful to compare
to a geodesic analogue of this behavior. Specifically,
we will play the same threshold game with geodesics
in a Kerr background by constructing a one parameter 
family $p$ of geodesics where for $p < \pstar$ the geodesics
eventually fall into the black hole, while for $p > \pstar$ they do not. 
For families of bound (elliptic) orbits, the latter subset of parameter
space exhibit {\em zoom-whirl} 
behavior---the geodesics start
some distance from the central black hole,
move in close to the black hole where they whirl around
for several orbits, then zoom out again to the original
distance; this behavior then repeats. For the
case $p<\pstar$ the initial behavior is similar, namely
the geodesic moves in from a distance and starts
whirling about the black hole, but then plunges into
the black hole. The number of whirl orbits increases
as $p$ approaches $\pstar$, going to infinity in the limit.
In this limit, the whirl part of the orbit comes arbitrarily
close to one of the unstable circular geodesics of the
spacetime (we are restricting attention to equatorial
orbits here). Qualitatively the same behavior is seen
in families of unbound (hyperbolic) geodesics, the only
difference is that there is only one episode of whirling
before the subset of geodesics that do not fall
into the black hole escape to infinity. 

This behavior is not only qualitatively similar to what is
observed in the full merger simulations, there is
also quantitative similarity comparing the scaling 
exponent $\gamma$ in (\ref{n_gamma}). Note
that there even {\em is} a geodesic analogue is surprising,
in particular if we take what the analogue is telling us
at face value: in this fine tuned regime the binaries
are approaching an {\em unstable circular} orbit. In the
full problem, one might guess that even if the binaries do temporarily 
approach what is an unstable circular orbit, then surely 
the ``perturbation''
implied by the rather strong gravitational wave emission
with this close separation would quickly force a merger? However
this does not seem to happen, and moreover 
we speculate that with sufficiently fine-tuned initial conditions
the binary can remain in this regime
until close to as much energy as is
theoretically possible is radiated away in gravitational waves
(more on this in Sec.\ref{sec_disc}).

In the remainder of this section we describe the
geodesic analogue in detail, first giving
an analytical calculation of
the scaling exponent $\gamma$ in Sec.\ref{geod_scale},
then showing
results from numerical integration in Sec.\ref{geod_int}.

\subsection{Calculating $\gamma$ from perturbation theory}\label{geod_scale}

Here we calculate $\gamma$ by finding the instability exponent $\lambda$
of unstable circular geodesics. Our analysis mirrors
the technique and formalism of \cite{CL} for Schwarzschild spacetime;
here we extend the result to equatorial orbits of Kerr.
We will perform the calculation in Boyer-Lindquist coordinates, 
in which the Kerr metric takes the following form
\begin{eqnarray}
ds^2=-\left(1-\frac{2 m r}{\Sigma}\right)dt^2 + \frac{\Sigma}{\Delta} dr^2 + 
     \Sigma d\theta^2 \nonumber \\
+R^2\sin^2\theta d\phi^2 - \frac{4 m a r\sin^2\theta}{\Sigma} d\phi dt,
\end{eqnarray}
where
\begin{eqnarray}
\Sigma &=& r^2 + a^2\cos^2\theta, \\
R^2 &=& r^2 + a^2 + \frac{2 m a^2 r \sin^2\theta}{\Sigma}, \\
\Delta &=& r^2 + a^2 -2 m r.
\end{eqnarray}
$m$ is the total mass of the black hole, and $J=a m$ the total angular momentum.

First, we want to relate $\lambda$, the growth rate of radial 
perturbations of the orbits in coordinate time $t$, to $\gamma$,
that characterizes how the number of orbits increases as 
the logarithmic distance to the critical parameter decreases.
The orbital angular frequency is 
\begin{equation}
\omega=\dot{\phi},
\end{equation}
where the overdot $(\dot{})$ denotes $d/dt$. 
We will start with a
geodesic with initial conditions $r(t=0)=r_0+\delta r_0$, where $r_0$ is the
radial coordinate of one of the unstable circular orbits,
and $\delta r_0$ is a small perturbation. We denote the
corresponding constant orbital frequency of the circular orbit with $\omega_0$, which is
\begin{equation}\label{omega_circ}
\omega_0=\frac{m}{ma\pm\sqrt{mr_0^3}},
\end{equation}
where $+$ sign is for co-rotating, the $-$ sign
for counter-rotating geodesics.
To linear order the growth of the perturbation $\delta r(t)\equiv r-r_0$
is given by
\begin{equation}
\delta r(t) = \delta r_0 e^{\lambda t}.
\end{equation}
Taking the absolute value and then natural logarithm of both sides we get
\begin{equation}\label{lndr}
\ln|\delta r(t)| = \ln|\delta r_0| + \lambda t.
\end{equation}
Perturbation theory breaks down when $\delta r(t) \approx 1$, which is
also the time when the geodesic will either leave the vicinity of the
black hole, or fall into it. By this time the number of orbits
that have been completed is $n=\omega_0 t/2\pi$. Finally, we note that
$\delta r_0$ will be proportional to $p-\pstar$ for a family of geodesics
that approach this unstable orbit when $p=\pstar$. Substituting all
this into (\ref{lndr}) and solving for $n$ gives
\begin{equation}
n = - \ln|p-\pstar| \frac{\omega_0}{2\pi\lambda} + {\rm const.},
\end{equation}
from which we can read off $\gamma$ (\ref{n_gamma}):
\begin{equation}\label{gamma_lambda}
\gamma=\frac{\omega_0}{2\pi\lambda}.
\end{equation}

Now we turn to the calculation of $\gamma$.
We restrict attention to equatorial ($\theta=\pi/2$) geodesics, for which 
the corresponding Lagrangian is
\begin{equation}
2\Screll = -\left(1-\frac{2 m}{r}\right)t^{'2} + \frac{r^2}{\Delta} r^{'2}  
+R_0^2 \phi^{'2} - \frac{4 m a }{r} \phi^{'} t^{'},
\end{equation}
where $R_0^2=r^2 + a^2(1+2m/r)$, and the prime ($'$) denotes differentiation
with respect to affine parameter $s$. The momentum $p_q$ conjugate
to geodesic coordinate $q^{'}$ is $\delta\Screll/\delta q^{'}$,
and in terms of $p_q$ and $q$ the Euler-Lagrange equations
for each pair $(p_q,q)$ is
\begin{equation}\label{EL}
\frac{dp_q}{ds}-\frac{\delta\Screll}{\delta q} =0 
\end{equation}
Given that $\Screll$ does not explicitly depend on $t$ and $\phi$ we 
immediately obtain the following two first integrals of motion,
which as usual we identify as the energy $E$ and angular momentum $L$
of the geodesics:
\begin{eqnarray}
E&\equiv& - p_t = \left(1-\frac{2 m}{r}\right)t^{'} + \frac{2 m a }{r} \phi^{'},\label{E_def} \\
L&\equiv&   p_\phi = R_0^2\phi^{'} -  \frac{2 m a }{r} t^{'} \label{L_def}.
\end{eqnarray}
A third constant of motion comes from the normalization condition
for timelike geodesics
\begin{equation}
-\left(1-\frac{2 m}{r}\right)t^{'2} + \frac{r^2}{\Delta} r^{'2}  
+R_0^2 \phi^{'2} - \frac{4 m a }{r} \phi^{'} t^{'} = -1.
\end{equation}
In principle we can use the above three equations to find a
first order differential equation for $r(s)$, though instead
we will follow \cite{CL} and evolve
the pair $(r,p_r)$ using (\ref{EL}). We shall also explore
the dynamics in terms of coordinate time $t$ rather than 
affine parameter $s$; the transformation between them 
can be obtained from (\ref{E_def},\ref{L_def}):
\begin{equation}
t^{'} = \frac{1}{\Delta} \left(E R_0^2 - 2maL/r\right)
\end{equation}
The resultant system of equations is
\begin{eqnarray}
\dot{r} &=& \frac{\Delta\ p_r}{r^2 t^{'}}\\
\dot{p}_r &=& \frac{t^{'} }{r^2}\left[\omega^2(r^3-ma^2)+m(2a\omega-1)\right] \nonumber\\
&+&\frac{p_r^2}{t^{'} r^3}\left[a^2-mr\right] \label{dot_pr_eqn}
\end{eqnarray}
Following \cite{CL}, we write the above as
\begin{equation}
\frac{dX_i(t)}{dt} = H_i(X_j),
\end{equation}
where $X_i(t)=(r(t),p_r(t))$ and $H_i(X_j)$ are the right hand sides of
the corresponding equations.
We linearize the equations about circular orbits, namely
let $X_i(t)=X_{i0}+\delta X_i(t)$ where $X_{i0}=(r=r_0,p_r=0)$, and only keep terms linear
in $\delta X_i(t)$:
\begin{equation}
\frac{d \delta X_i(t)}{dt} = K_{ij} \delta X_j(t),
\end{equation}
where
\begin{equation}\label{kdef}
K_{ij} = \frac{\partial H_i(X_j)}{\partial X_j}|_{X_i=X_{i0}}.
\end{equation}
The quantity $\lambda$ that we are interested in is
the positive real eigenvalue of (\ref{kdef}), which exists
for the range of $r_0$ corresponding to unstable circular
orbits. We can then substitute this into (\ref{gamma_lambda})
to find $\gamma$. After a tedious but straight-forward calculation we get
\begin{equation}\label{gamma_res}
\gamma=\frac{r^2}{2\pi} \left [3r^2\Delta + \frac{4m}{\omega^2}
\left(rR_0^2\omega^2 - 4ma\omega -r + 2m\right)\right]^{-1/2},
\end{equation}
where in the above $\omega$ and $r$ are evaluated at $\omega_0$
and $r_0$ respectively.
Note that although the preceding calculation is gauge dependent, the final
result (\ref{gamma_res}) only directly refers to the radial 
coordinate $r$, and indirectly to the azimuthal coordinate $\phi$ 
in that we measure the number of orbits $n$ completed by the geodesic.
$r$ can be eliminated entirely from the
expression in favor of the proper circumference $2\pi R_0$
of the geodesic (though for simplicity we have not done so),
and given the axisymmetry of the spacetime one can unambiguously define
$n$.

\subsection{Equatorial geodesic orbits in Kerr}\label{geod_int}

In the preceding section we calculated $\gamma$ by examining
the growth of a perturbation of an unstable circular geodesic.
However, it might not be obvious that one parameter
families of geodesics that smoothly interpolate between capture
and non-capture will approach one of these unstable orbits
as the limiting case between capture and non-capture.
Here we show a few examples that this is indeed the generic
behavior at threshold for equatorial geodesic families, though the
{\em particular} unstable circular orbit that is approached depends on the
initial conditions.

\begin{figure}
\begin{center}
\includegraphics[width=8.5cm,clip=true]{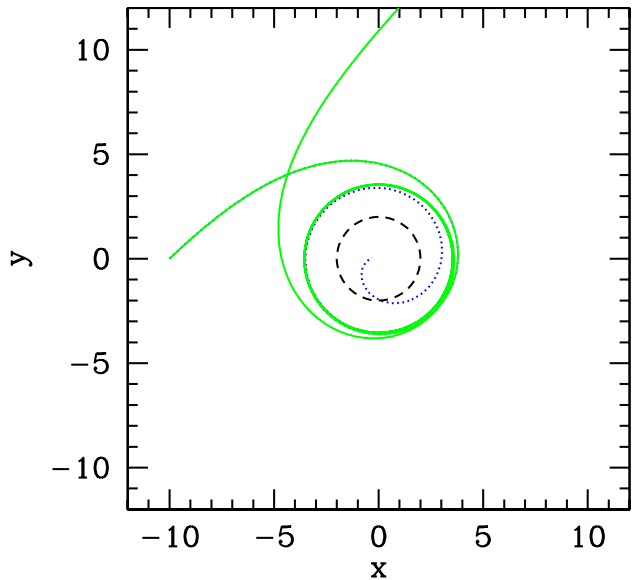}
\end{center}
\caption{
Plots of two timelike geodesics of Schwarzschild ($m=1$) spacetime tuned
close to the threshold of capture. The geodesic integration
was started at $(x,y,z)=(-10,0,0)$, the geodesics given a velocity
of $v$, and the parameter used to find the threshold was
the initial angle $\theta$ of the tangent vector relative
to the x-axis. Here, the two geodesics evolve toward an unstable
circular orbit at $3.54$, remain close to it for roughly 6 orbits,
then the one with $\theta<\theta^*$ (dotted curve) falls into the black hole
while the other (solid curve) escapes. The dashed circle
is the location of the event horizon. Note that the unstable
circular orbit that is approached has $r<4$, which is consistent
with the orbit being unbound (see for example \cite{MTW_orbit}).
}
\label{fig_d_sbh1}
\end{figure}

\begin{figure}
\begin{center}
\includegraphics[width=8.5cm,clip=true]{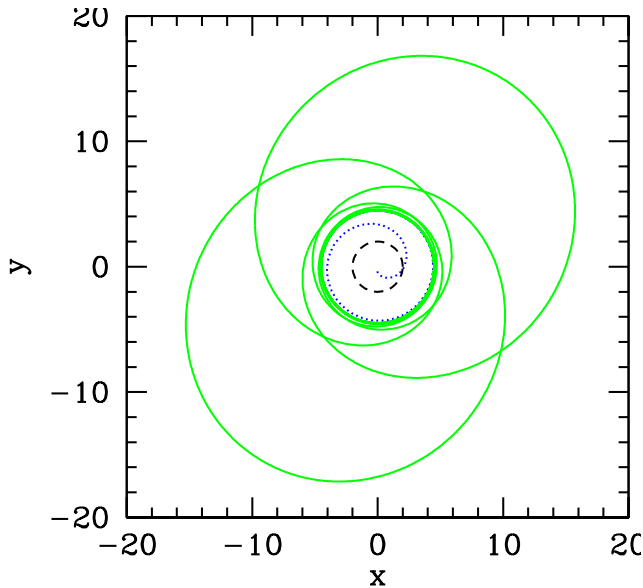}
\end{center}
\caption{
A second example of timelike geodesics of 
Schwarzschild ($m=1$) spacetime tuned
close to the threshold of capture. Here, the geodesic integration
was started at $(x,y,z)=(-4.5,0,0)$, and the parameter
tuned to threshold was the initial velocity $v=(0,v_y,0)$ in the
$y$ direction. In other words, we are effectively starting on an unstable
circular orbit with a ``perturbation'' $v_y-v_y^*$.
The two geodesic orbits shown here remain close to $r=4.5$ 
for about 8 orbits, one (dotted curve) then falls into the
black hole, while the other escapes. Since $r>4$ this is a bound
orbit (see for example \cite{MTW_orbit}), and the geodesic ``zooms'' out to 
some distance before
returning to $r\approx4.5$ to undergo another ``whirl'' episode.
Here we show two full zoom-whirl episodes (and note that
this is {\em not} a closed orbit; the integration was stopped
during the start of the third whirl phase).
}
\label{fig_d_sbh3}
\end{figure}

\begin{figure}
\begin{center}
\includegraphics[width=8.5cm,clip=true]{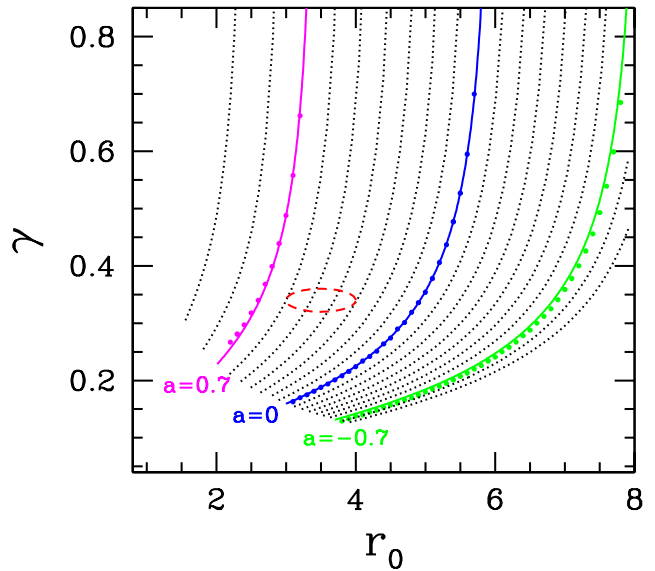}
\end{center}
\caption{Plots of $\gamma(r_0)$ for Kerr equatorial 
geodesics (\ref{gamma_res}), for values of the spin
parameter $a$ ranging from $a=-0.9$ (right-most curve)
to $a=0.9$ (left-most curve) with intervals of $0.1$. 
For three values of $a$ ($0,\pm0.7)$ we plot $\gamma(r_0)$
calculated using geodesic integration tuning to the
threshold of immediate capture. Specifically, {\em each point}
depicted in the figure shows results of a bisection
search to the threshold of capture for a one parameter
family of geodesics; the value of $r_0$ is extracted
from the nearest-to-threshold solution, and $\gamma$ is
obtained from data similar to that shown in Fig.\ref{fig_gamma_merge}.
The dashed ellipse shown in the figure is the value
of $\gamma$ (enlarged from a point to include estimated
uncertainties) extracted from the fully non-linear evolutions
described in Sec.\ref{sec_threshold}. This is in rather
remarkable agreement with the geodesic results given that the
final spin of the black hole is $0.68\pm0.05$, and the trend
from the simulations shows $r_0$ is slowly decreasing
as one approaches the threshold, suggesting that the ellipse
would shift to the left if we could tune closer to threshold.
The near-critical geodesics at the center of the 
dashed ellipse have an eccentricity of $e\approx0.5$, which might
be one way to define what the effective initial eccentricity
of the equal mass system is.
%
%
%
%
%
}
\label{gamma_rmin_a_geod}
\end{figure}

Figs. \ref{fig_d_sbh1} and \ref{fig_d_sbh3} show a couple of examples
for two fine-tuned families of initial data in Schwarzschild spacetime
($a=0$), integrated using the method explained in Appendix \ref{sec_geod_details}
(and note that we have integrated the geodesics in Cartesian coordinates
in the Kerr-Schild form of the metric, which is horizon penetrating).
Fig. \ref{fig_d_sbh1} is an example of a hyperbolic family of orbits, 
Fig. \ref{fig_d_sbh3} an elliptic family. This again illustrates
that no special care need be taken in choosing a class of orbits
to exhibit this behavior, merely that the orbits can be labeled
by a parameter that smoothly intersects the threshold of capture.
Fig. \ref{gamma_rmin_a_geod} shows $\gamma$ as a function
of $r_{0}$, the radius of the geodesic in the whirl-phase, for a
range of values of the spin parameter $a$. The lines in the figure
were calculated using (\ref{gamma_res}), and for three representative
cases of $a$ we overlay the results from a numerical calculation
based on the same method used to calculate $\gamma$ for the fully non-linear
problem described in Sec.\ref{sec_threshold}. Note that each
point from the numerical calculation represents fine-tuning a family
of geodesics to the threshold of capture, and we typically tuned
the initial conditions to within 1 part in $10^{12}$ (being geodesic
integration on a fixed background this is a ``cheap'' problem). 
It was also not difficult to find sets of one parameter families
that at threshold spanned most of the range of unstable circular
orbits. For example, repeating the exact numerical setup of
the binary black hole problem, we can choose the geodesic to
have some initial coordinate $(x,y,z)=(d,0,0)$ and an initial
velocity $(0,v_y,0)$. Keeping $d$ fixed and tuning $v_y$ to threshold
we approach an unstable circular orbit the radius of which depends
on $d$. So repeating this threshold search for a range of $d$ we
can map out most of the range of unstable circular orbits.

On Fig. \ref{gamma_rmin_a_geod} we also plot an ellipse
representing $\gamma$ and $r_0$ measured in the equal mass
problem described in Sec.\ref{sec_threshold}; the size
of the ellipse depicts the numerical uncertainty in this
``point''.  For $r_0$ we have used the coordinate
separation between the black holes, which, based on how
well this quantity reproduces the gravitational waves
emitted when plugged into the quadrupole formula 
(see Fig.\ref{fig_Lm2_c22r}) suggests
this is a reasonable distance indicator not subject to severe gauge
artifacts\footnote{Another suggestion might be to
use the proper distance between the two horizons;
however this, when used in the quadrupole formula, overestimates
the energy by almost a factor of 2.}. 
The region of the diagram where $\gamma$ and $r_0$ from
the binary merger simulations fall lends support to the idea
that the geodesic analogue {\em is} describing what is happening in the
full problem---the co-rotating geodesic orbit 
with the same value of ($r_0$,$\gamma$) as the non-linear
problem gives a spin parameter of $\approx 0.5$, which is not too far from
that of the final merged black hole of $\approx 0.7$.

\section{Discussion}\label{sec_disc}

In Sec.\ref{sec_threshold} we gave evidence of a ``threshold
of immediate merger'' in the binary black hole problem
for one parameter families of initial data that 
interpolate smoothly between a scenario of prompt
merger at one extreme and non-merger (or merger much
further into the future) at the other extreme. Near threshold
evolutions were marked by a period of near-circular orbital evolution 
at very close separation, significant gravitational wave
emission, and exponential sensitivity to initial conditions.
In Sec.\ref{sec_geod} we demonstrated that this behavior
was quantitatively similar to equatorial geodesic behavior in a 
Kerr background, where the corresponding threshold is that
of capture of the geodesic. Without the geodesic analogue
this behavior in the full problem might seem very bizarre,
and indeed even with the geodesic analogy it is still rather
surprising. For given how unstable the circular orbits are
that are approached at threshold, one would imagine that 
a ``perturbation'' such as the emission of a significant
amount of gravitational radiation per orbit would make
such an orbit unattainable away from the geodesic approximation.
However, as we will try to argue in Sec.\ref{sec_gen} below, 
this behavior is 
almost ``obvious'' if we try to imagine the possible
orbits that could arise in the full problem. The argument
applies to the generic scenario of unequal mass, arbitrary spin
initial conditions, however the argument cannot claim that generically
there will be any relationship with unstable geodesics
of black hole spacetimes. Near-threshold orbits almost
certainly do {\em not} have astrophysical significance due
to the fine-tuning required to reach them, however 
if large extra dimensions exists and will result in 
significant black hole production at the LHC, this behavior could
have some application there---we discuss this in Sec. \ref{sec_lhc}.

\subsection{Beyond equal mass non-spinning mergers}\label{sec_gen}

\begin{figure}
\begin{center}
\includegraphics[width=7.3cm,clip=true]{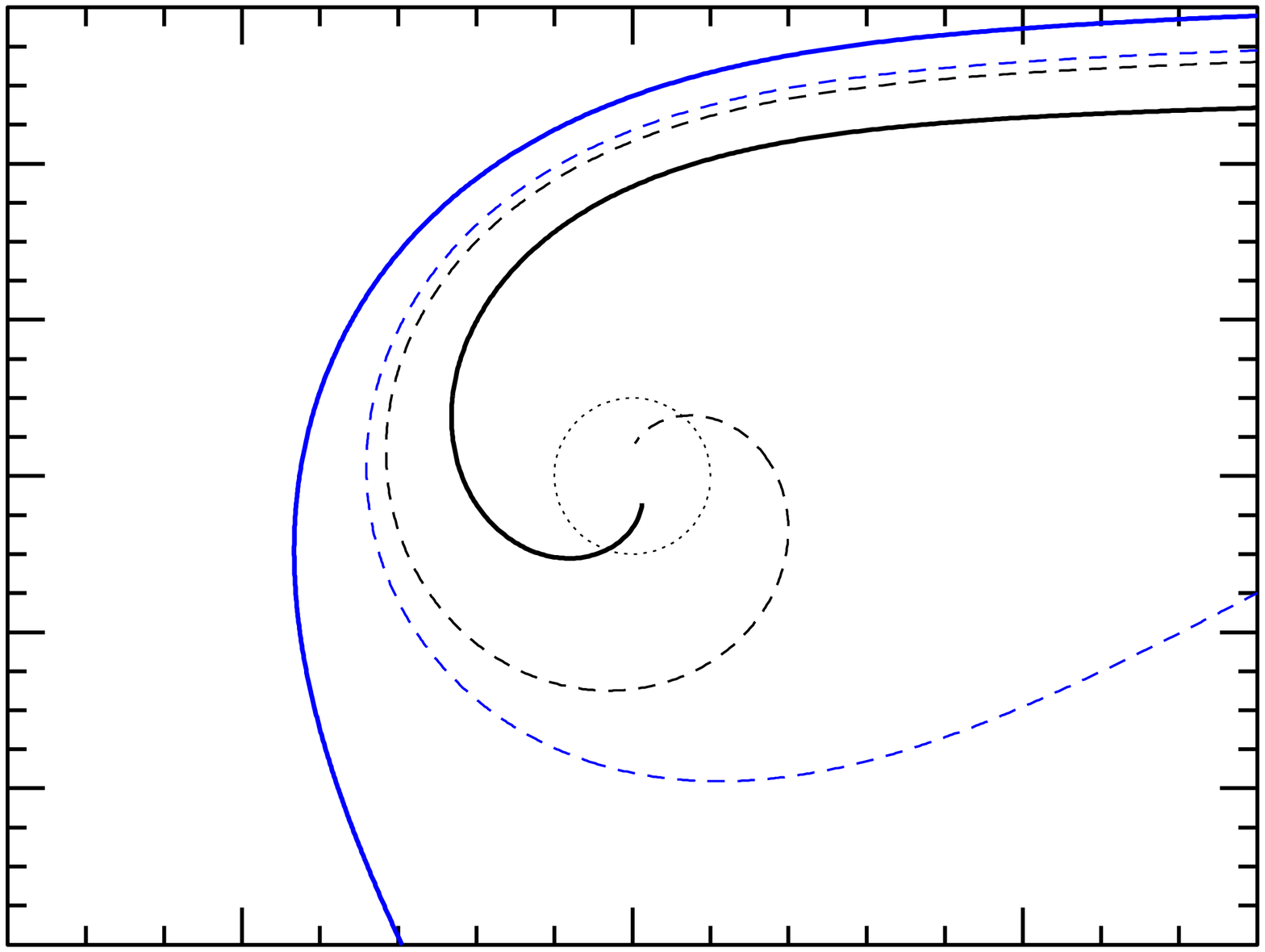}
\includegraphics[width=7.3cm,clip=true]{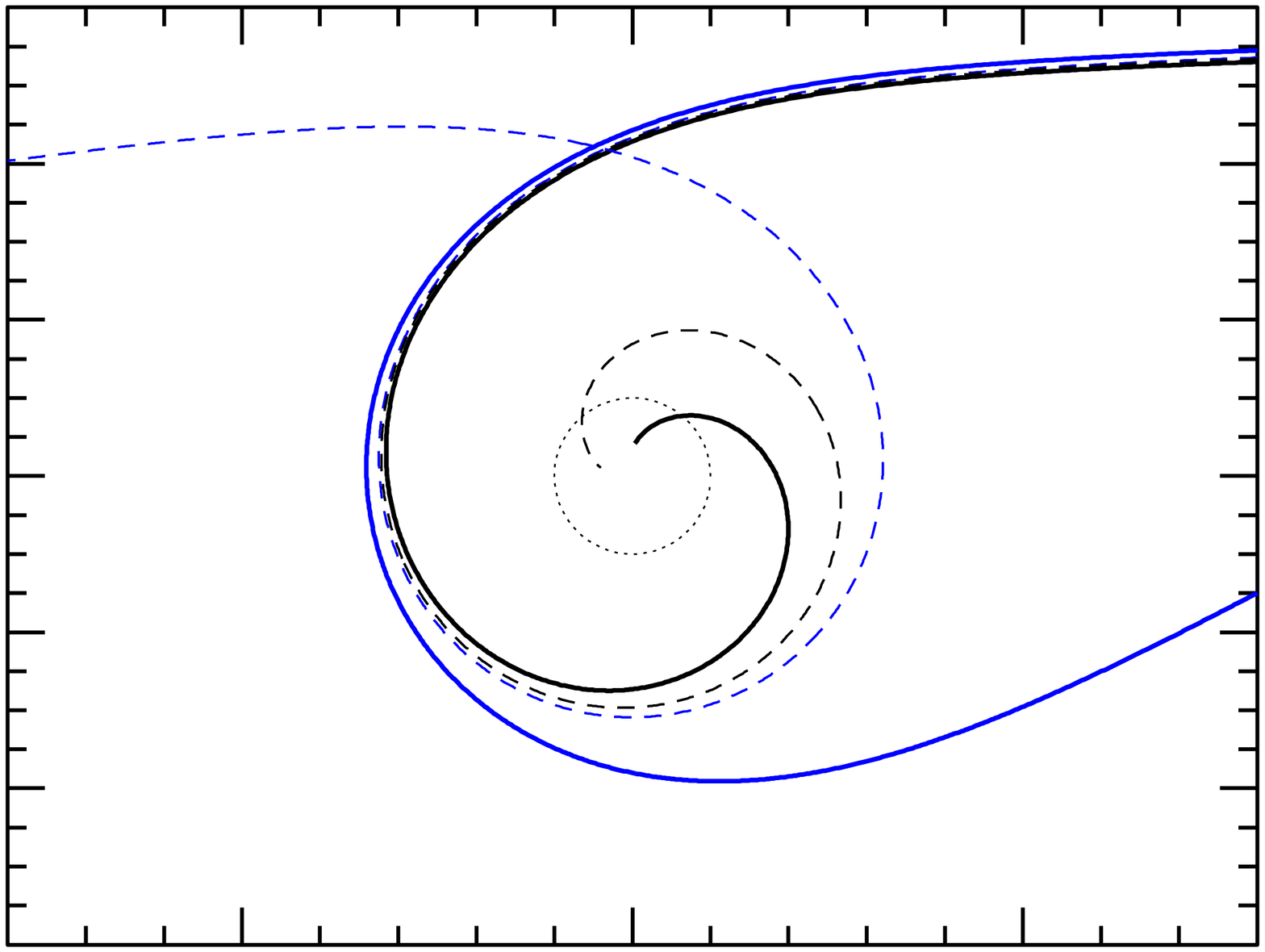}
\includegraphics[width=7.3cm,clip=true]{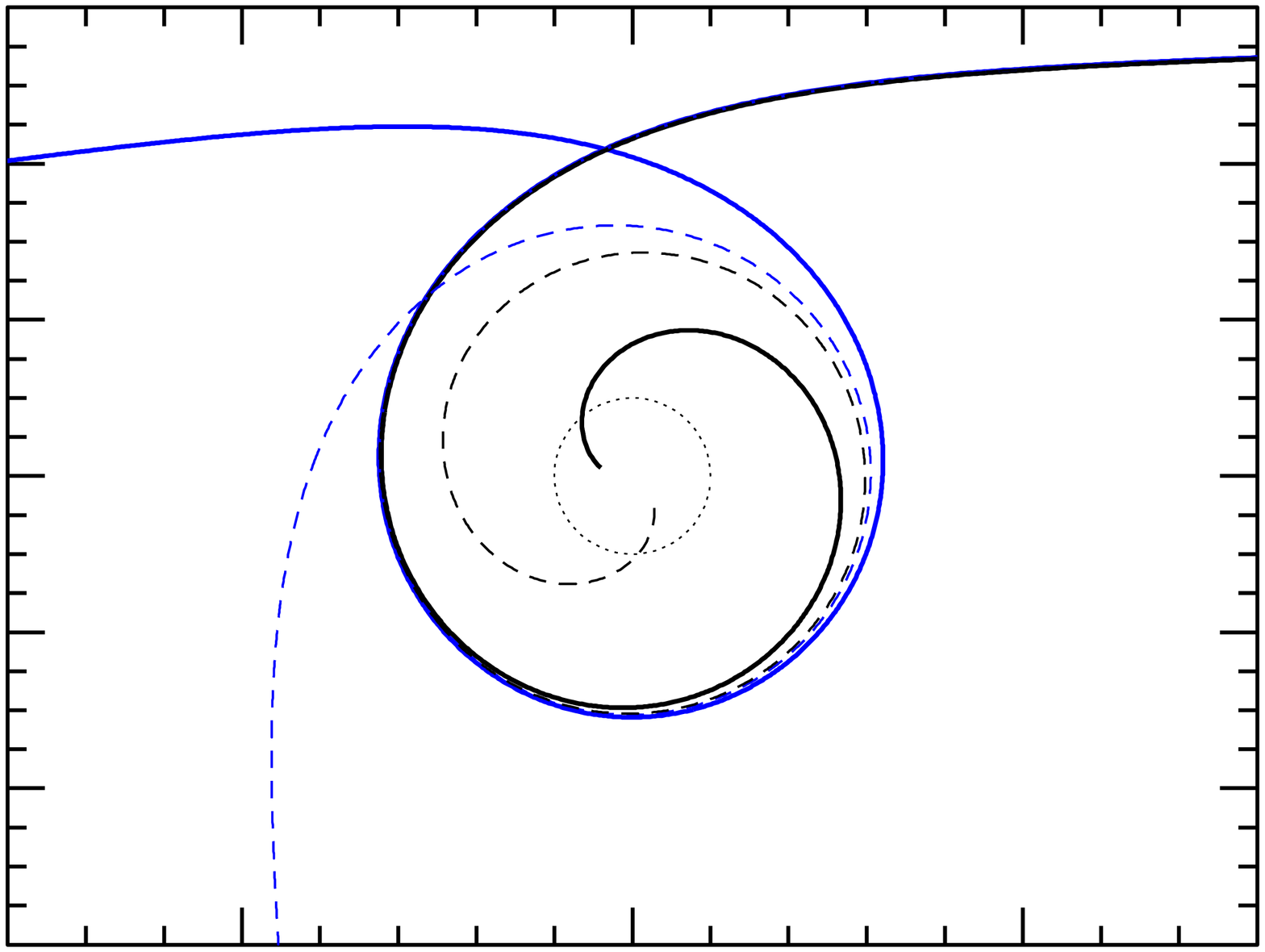}
\end{center}
\caption{A schematic illustration of why an immediate
threshold must exist for the generic scattering problem.
Each panel shows several hypothetical trajectories
of one black hole in a two body interaction in the center
of mass frame; for clarity the second black hole is not
shown.
The two thicker solid curves represent impact parameters
bracketing the threshold, i.e. in one case there is a merger,
in the other a deflection. The two thinner dashed curves
represent the two possible trajectories for an impact
parameter between the bracketing cases. Going from the
top to bottom panel we are successively tuning closer to threshold.
It is difficult to imagine how a bisecting trajectory
could {\em not} follow a path {\em between} the two bracketing
trajectories as illustrated, and so tuning to
threshold one forces the binaries into a close
``whirling'' configuration.
}
\label{fig_it_ex}
\end{figure}

Consider a smooth one-parameter family $b$ of initial conditions 
for two black holes of arbitrary masses and spins that straddle
a regime of immediate merger.
We further assume that the parameter is monotonic in that
the threshold is only crossed once as $b$ is varied (this also
assumes that the threshold is not fractal in nature).
Let $b=b_1$ denote one extreme (merger), $b=b_2$ the other (deflection).
Now choose a value of $b$ halfway between these two: $b_3=(b_1+b_2)/2$. 
By the assumed smoothness and monotonicity of $b$ the resultant
orbit must lie {\em between} the first two 
orbits\footnote{Though with one or both of the black holes
possessing spin the orbital plane will in general not lie on
a plane, making the trajectories curves in 3D space and
preventing a simple and precise notion of ``between''.},
and given that one orbit is anchored in a merged black hole, the
black holes in the new orbit must spend more time in
close proximity than either of the preceding cases
before merging or separating---this is
illustrated in Fig. \ref{fig_it_ex}. Continuing
the bisection therefore forces the corresponding
binary system to approach an every lengthier whirl-like 
period of evolution.
However, since more time is spent in a tight
orbit, more energy will be radiated, and if the binary
was initially unbound, after a certain amount of fine-tuning
the system will become bound. Continuing the process
beyond this point (or from initial conditions that were
bound to begin with), the effective
apoapsis of the orbit for the cases that separate
must decrease as the threshold is approached. Gravitational
radiation therefore puts a limit to this process, ending
it once enough energy has been radiated so that the next apoapsis
decreases to the radius of the orbit the binary
is on during the whirl phase. Of course,
approaching this point the notion of an ``immediate merger''
will become ill-defined, for ever sooner after separating the binaries will
merge and thus it will become impossible to differentiate
between prompt merger and separation.

The preceding argument appeals to continuity of the orbital
trajectories as a function of a smooth initial data parameter $b$. One
way of preserving continuity, yet having the whirl regime
terminate before all possible energy has been
radiated, is if at some distance from threshold the encompassing
black hole forms while the two black holes are still in the whirl
phase. However, given that
both the simulations shown in Sec.\ref{sec_threshold} and
the geodesics in Sec.\ref{sec_geod} have a distinct
plunge phase for the cases that do merge, we think this
possibility is unlikely. One interesting question then
is how much binding energy could be extracted at threshold
via gravitational waves. If cosmic censorship
holds then an upper limit is given by Hawking's area
theorem. The exact amount depends on the angular momentum
of the final black hole; if it is non-rotating a maximum
of $29\%$ of the rest mass could be radiated; 
for the case studied here if the trend 
continues and the final spin remains $a\approx 0.7$,
$24\%$ could be radiated. 
In the high-speed scattering problem where the
kinetic energy of the black holes dominate the net energy,
if the argument outlined in the previous paragraph holds
then it may be possible for {\em all} the {\em kinetic} energy of the
system to be emitted as gravitational waves at threshold. 
This can be an arbitrarily large fraction of the total
energy of the system. 

We note that for the above arguments to be valid
it is {\em crucial} that two distinct end-states exist in the
scattering problem. In Newtonian gravity for two point masses, 
for example, this type of behavior cannot exist because 
there is no merger end-state. It is also interesting that
in the geodesic problem the whirl-type behavior that arises
at threshold is intimately connected to the existence
of unstable circular orbits; this makes it tempting to think
that there may be some deep connection between the existence
of unstable geodesics in Kerr and the fact that black holes {\em merge}
when they collide in general relativity.

\subsection{Black hole scattering at the LHC}\label{sec_lhc}

One interesting application of this apparent analogy
between the threshold of capture of geodesics and
the threshold of immediate merger in the full problem
is to obtain some understanding of the black hole scattering
problem. This is of interest in the search for black
holes that might be formed at the LHC (for a few review articles
see \cite{landsberg,kanti,gingrich}). Consider the 
scattering of two black holes colliding with an impact parameter $b$. 
We would like to know $E_r(b)$, the energy radiated
as a function of the impact parameter $b$, and the threshold
impact parameter $\bstar$ below which a merger results
(this would correspond to the threshold of black hole production
in a particle collision of sufficiently high energy where the
classical general relativistic description of the event holds).
Below we illustrate how to find an approximation to $E_r(b)$
in the 4 dimensional case; certainly to be of relevance to
the LHC this calculation will need to be carried out in 
higher dimensions, and many of the approximations made
could be improved upon, though here we are more interested 
in presenting the main ideas. Note that higher dimensional
black hole spacetimes in general do not posses stable circular
orbits\cite{tangherlini,frolov_stojkovic,cardoso_yoshida}, 
though what is crucial here is the existence of {\em unstable}
circular orbits.

For concreteness we consider the interaction of two equal
mass, non spinning black holes of total rest mass $m$
traveling with large center-of-mass speeds $v$.
In this regime the total energy of the
system $E=\Gamma m \equiv m/\sqrt{1-v^2}$ is dominated 
by the kinetic energy of the black holes. Also, for
sufficiently large $\Gamma$ any spins or electric
charges of the black holes will become irrelevant, 
as these quantities are Lorentz invariant and thus
the ratio of spin-to-kinetic and/or 
electromagnetic-to-kinetic energy of the system goes to zero.
So the non-spinning, uncharged case will be a good approximation
to the generic high energy scattering problem\footnote{Though at energies
probed by the LHC, and given that current experiments
suggest the Planck scale cannot be very far below
maximum LHC energies, the effects of spin and
charge would be important to consider\cite{gingrich2}.}. Denote
the threshold of immediate merger by $b=\bstar$.
Guided by the results presented in earlier sections, we
will assume the following key pieces of information to obtain an
approximation to $E_r(b)$:
\begin{itemize}
\item evolutions near the threshold of immediate merger $\bstar$
      are characterized by the scaling relation (\ref{n_gamma}):
      $e^n\propto|b-\bstar|^\gamma$
\item the critical impact parameter $\bstar$ and the scaling exponent
      $\gamma$ can be found by considering
      the analogous problem of geodesic scattering in a Kerr background. The
      ADM mass of the background space time is $E$ (which sets the
      scale for $b$), and has a spin parameter $a$ 
      equal to that of the black hole that forms in the full 
      problem for $b$ slightly less than $\bstar$. 
\item during the whirl phase, a constant fraction $\epsilon$ of the remaining
      energy $E-E_r(t)$ of the system is radiated per orbit
\end{itemize}
There are a couple of additional bits of information that will be needed
to complete the calculation---the energy $E_{r0}\equiv E_r(0)$ emitted in
a head-on collision, the value of $a$ for the background
Kerr spacetime, and the fraction $\epsilon$ of 
energy radiated per orbit. We will use an existing estimate
of $E_{r0}$ from the literature, argue below that $a$ slightly
less than extremal ($a=1$) is the relevant parameter in the
large $\Gamma$ limit, and use the quadrupole formula
for $\epsilon$. Using the quadrupole formula together
with geodesic motion is similar in spirit to the
``semirelativistic approximation'' used to compute
waveforms for extreme mass ratio inspiral \cite{ruffini_sasaki,gkl}.

To simplify the calculation we will use a normalized
energy $\Ebar\equiv E_r/E$, and define the following normalized
impact parameter:
\bea
\bbar&\equiv&\frac{b}{\bstar}, \ \ \ 0 \le b \le \bstar \\
\bbar&\equiv&\frac{b+b_c-2\bstar}{b_c-\bstar}, \ \ \ \bstar \le b \le b_c. 
\eea
$b_c$ is a cut-off value for the impact parameter beyond
which (\ref{n_gamma}) ceases to offer a good approximation to the scattering
behavior; geodesic integrations suggest $b_c\approx2\bstar$ in most cases.
With the above normalization $0\le\bbar\le2$, and 
(\ref{n_gamma}) becomes
\be\label{bbg}
n(\bbar)= - \gamma \ln |1-\bbar|.
\ee
In the above we have added the additional approximation
that $n(\bbar=2)=0$; strictly speaking $n$ is only
zero at $b=0$ (for non-spinning black holes) and in the limit 
as $b\rightarrow\infty$.
We assume that the energy $\Ebar$ emitted during the process
is simply a function of $\bbar$, and hence $n$ by
the above relation
\be\label{dedB}
d\Ebar/d\bbar = \frac{d\Ebar}{dn} \frac{dn}{d\bbar},
\ee
and that a constant fraction $\epsilon$ of the remaining
energy is emitted per orbit in the whirl phase:
\be\label{dEdn}
d\Ebar/dn = \epsilon (1-\Ebar)
\ee
Integrating (\ref{dedB}) with (\ref{bbg},\ref{dEdn}) then gives
\bea
\Ebar(\bbar) &=& 1 - \left(1-\Ebar_0\right)
     \left(1-\bbar\right)^{\epsilon\gamma}, 
     \ \ \ 0\le\bbar\le1 \nonumber \\
\Ebar(\bbar) &=& 1-(\bbar-1)^{\epsilon\gamma},
\ \ \ 1\le \bbar\le 2\label{eb}
\eea
For boundary conditions we have assumed all the energy is
radiated for $\bbar=1$, $\Ebar_0$ is the
energy radiated for the head-on collision case, and
$E$ drops to zero at $\bbar=2$. 

\begin{figure}
\begin{center}
\includegraphics[width=8.5cm,clip=true]{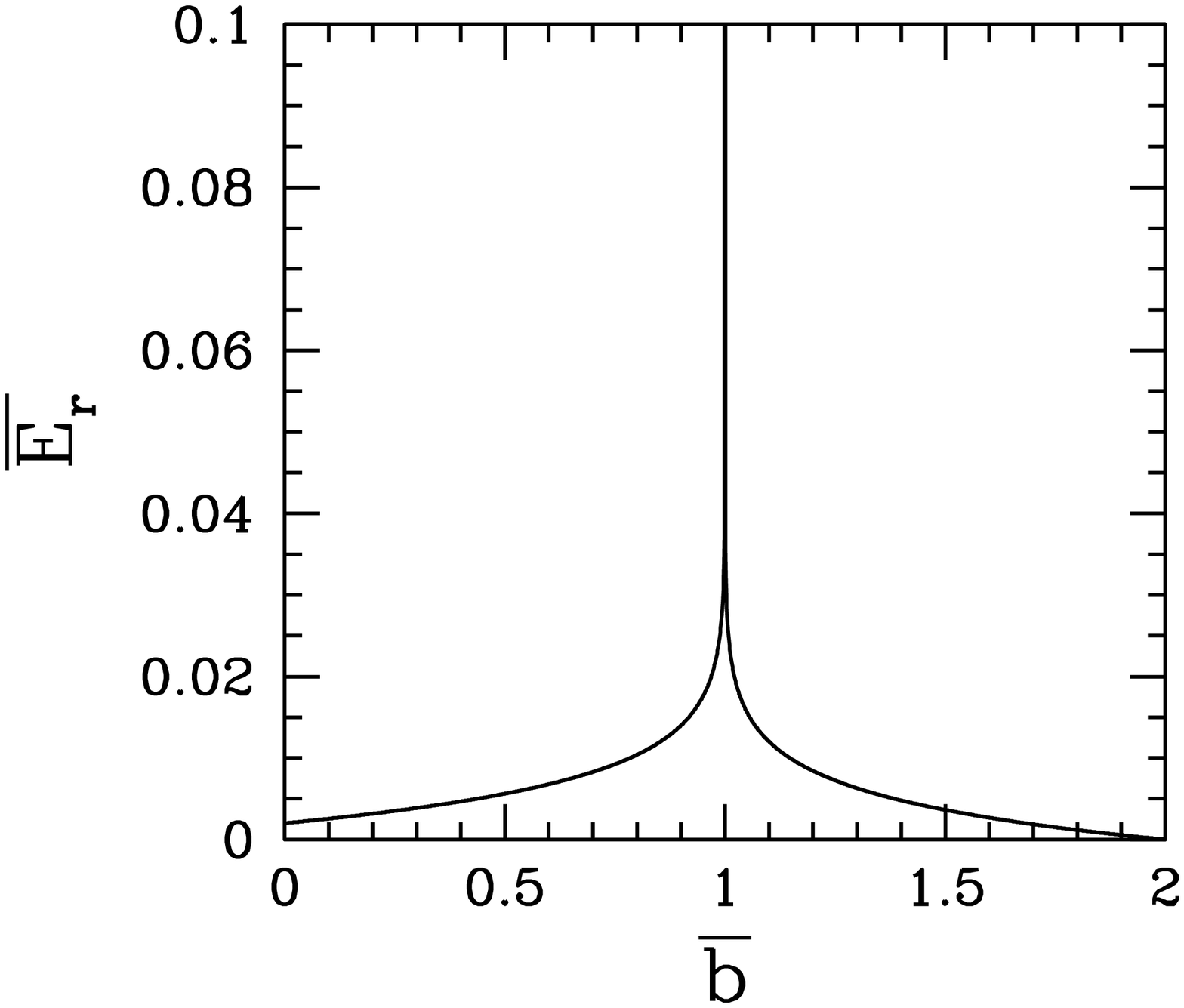}
\includegraphics[width=8.5cm,clip=true]{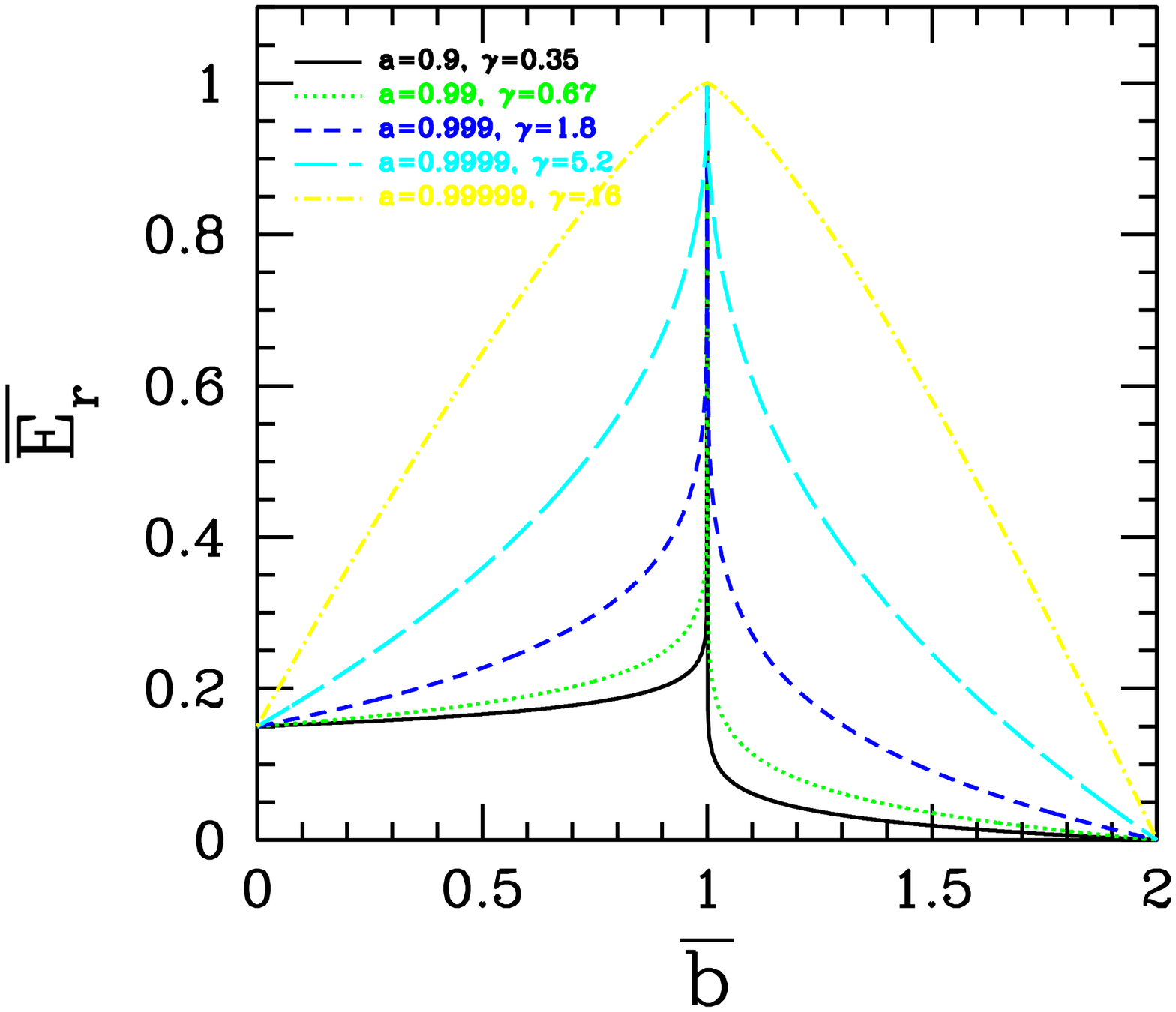}
\end{center}
\caption{Estimated fraction of the total energy radiated as
a function of the impact parameter for the black hole
scattering problem (\ref{eb}). The top panel illustrates
a rest-mass dominated case, using parameters from the simulation
results presented in Sec.\ref{sec_threshold}; here cosmic censorship
would place an upper bound for $\Ebar(1)$ of $\approx 0.29$.
The bottom panel illustrates several curves from the kinetic
energy dominated regime (note the different vertical scales in the
two panels). For a concrete example we have used the 
value of $0.15$ for $\Ebar_0$, which is about half that of the
upper limit computed using the trapped surface 
method (for a review of various methods see \cite{cardoso_et_al}).
The order-of-magnitude calculation described in the text suggests
that in the ultra relativistic limit the final
state will be an extremal Kerr black hole (with negligible
mass relative to the initial kinetic energy of the system). 
In this limit the geodesic analogue calculation
becomes very sensitive to the value of the final
spin parameter $a$, and so we show several curves for 
different values of $a$. In the limit $a\rightarrow 1$,
$\gamma\rightarrow\infty$, and the estimate $\Ebar(\bbar)$ 
in (\ref{eb}) approaches a unit step function $\Theta(\bbar/2)$.
}
\label{fig_b_E}
\end{figure}

For interest, in the top panel of Fig.\ref{fig_b_E} we show a plot of (\ref{eb})
with parameters plugged in from the low-speed system
discussed in Sec.\ref{sec_threshold}. Of course, in this
case $\Ebar(1)$ can at most be $\approx.29$ from
the area theorem, though for values of $\bbar$ such that $\Ebar(\bbar)<\approx.29$
(\ref{eb}) is still a very good approximation.

\begin{figure}
\begin{center}
\includegraphics[width=8.5cm,clip=true]{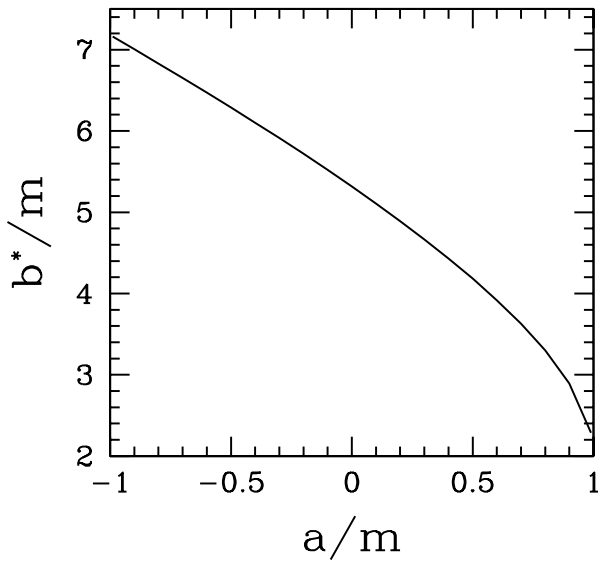}
\end{center}
\caption{The critical impact parameter $\bstar$ for
equatorial geodesics with initial Lorentz $\Gamma=10$
on a Kerr background with spin
parameter $a$ and mass $m$.
}
\label{b_star_a_gamma_10}
\end{figure}

Returning to the ultra-relativistic problem, we now
estimate $\epsilon$ using the quadrupole
formula. For a circular
orbit composed of equal mass point particles
separated by a distance $\bar{r}$ and orbiting with angular
velocity $\bar{\omega}$, where the over-bars ($\bar{\ }$) denote
that the quantities have been normalized to the remaining
energy $1-\Ebar$ in the system, the quadrupole formula gives
\be
\epsilon_{quad}=\frac{4\pi}{5}\bar{r}^4\bar{\omega}^5.
\ee
So what values of $\bar{r},\bar{\omega}$ and $\gamma$ to use?
In the ultra-relativistic geodesic scattering problem, the geodesics approach
the light-ring (unstable circular {\em photon} geodesic) of the
background spacetime at threshold. It also seems natural to
guess that in this limit in the full problem the final angular
momentum of the black hole will approach extremality. The initial
angular momentum of the system with critical impact
parameter is $J=\bstar E/2$ (restoring units); thus the initial
{\em effective Kerr parameter} of the orbit is $a_o=\bstar/2$. 
A plot of $\bstar$ versus the background Kerr parameter
$a$ for geodesic motion is shown in Fig.\ref{b_star_a_gamma_10}---note that
in the limit $a/M\rightarrow 1$, $a_0\rightarrow 1$. To estimate
how $a_o$ changes during evolution, we again use quadrupole
physics, which says $dJ/dt = \omega^{-1} dE/dt$, and define
the instantaneous $a_o(t)\equiv J(t)/E(t)$. We then get
\be
\frac{d(a_0/E)}{dn} = \frac{dE}{dn}\frac{1}{E}\left(\frac{1}{E\omega} 
-2\frac{J}{E^2}\right)
\ee
In the limit $a/M\rightarrow 1$, $J/E^2\rightarrow 1$ initially,
and using the Boyer-Lindquist value (\ref{omega_circ}) for $\omega$ for a
photon on the light-ring in the extremal case, we 
get $E\omega\rightarrow 1/2$. In other words, $d(a_0/E)/dn=0$ in this
limit, so at least to within the quadrupole approximation
assuming an extremal Kerr background for the geodesic analog
in the ultra relativistic limit is self consistent.

Though now we have a bit of a dilemma---in the extremal limit
there are {\em no} unstable circular orbits of Kerr, and hence
$\gamma\rightarrow\infty$! In a practical setting (such as the LHC)
one of course will not be at the limit, though given how sensitive
$\gamma$ is to $a$ approaching the limit it will be difficult
to justify any crude estimates such as that outlined
in the previous paragraph to decide which
value of $a$ to use to determine $\bstar$ and $\gamma$
from geodesic motion. In the bottom panel of Fig.\ref{fig_b_E} we therefore plot
several possibilities for $\Ebar(\bbar)$ with a few values
of $a$ close to 1.  We have used the limiting value of 
$\epsilon_{quad}\rightarrow\pi/40$, which is less sensitive
to $a$ than $\gamma$ in the limit.

If Fig.\ref{fig_b_E} gives a decent description of the ultra relativistic 
particle scattering problem, then even though the cross section for 
black hole production 
will to a good approximation be $\pi\bstar^2/4$, there can still be significant energy
loss to gravitational waves for $b$ up to almost twice $\bstar$ (or 
possibly even more, since recall that $\Ebar(\bbar=2)=0$ was only 
an approximate boundary condition we imposed).
$\bstar\approx 2 E$
in the limit, which implies an effective cross section for 
measurable energy loss of
$4\pi E^2$.
It is interesting that this number
is identical to the order of magnitude estimate made by 
assuming the cross section is equal to $\pi R_s^2$, where 
$R_s=2E$ is the Schwarzschild radius corresponding to the initial
center-of-mass energy. We finally note that a lower limit on the impact
parameter resulting in black hole formation can be computed
by searching for trapped surfaces at the moment the two
shock waves representing the $\Gamma\rightarrow\infty$ particles
collide; in \cite{yoshino_rychkov} trapped surfaces were found
for $b<\approx 1.68 E$.

\section{Summary and Conclusions}\label{sec_conclude}

In this paper we have described numerical simulations of the
merger of a class of equal mass, zero initial spin, non-circular 
binary black hole systems in general relativity. For a one
parameter ($k$) family of solutions interpolating between a deflection
of the binaries without merger at one extreme of the parameter, and 
a prompt merger at the other extreme, we provided evidence that
there is a notion of a threshold of immediate merger at $k\approx\kstar$
during with the binary enters a tight near-circular whirl configuration
before either separating or merging. The number of orbits $n$
spent in the whirl is exponentially sensitive
to the initial conditions: $e^n\propto |k-\kstar|^\gamma$, where
$\gamma$ is approximately a constant. The area theorem
together with measurements of the energy lost to gravitational
waves suggests this whirl behavior could persist for as many as
20-30 orbits for extremely fine tuned initial data, though
we have only been able to tune to $\approx 5$ orbits due
to limited computational resources.

A second result of this work has been to show that 
similar behavior is observed in the analogous problem of 
the scattering of a one parameter family of equatorial geodesics off a Kerr black
hole. At threshold, the geodesic approaches one of the unstable
circular orbits of Kerr at a radius $r_0$ that depends on the
particular family of geodesics. One quantitative similarity
between the test particle and equal mass cases 
is we notice roughly the same scaling exponent $\gamma$ 
with geodesics orbiting a black hole with
spin parameter $a$ close to that of the black hole
that forms in the merger case in the full problem, and
approaching an unstable orbit with radius similar to the
coordinate separation of the black holes during their
whirl phase. Comparison with the quadrupole formula for the
emitted gravitational waves suggests the simulation coordinates
are well adapted to the underlying physics. 

The close analogy between threshold geodesic scattering and the one
class of equal mass interactions studied here, together with
an argument that such a threshold should exist for generic
one parameter families of initial configurations with
appropriate limiting cases, gives us some confidence in
extrapolating this behavior to the kinetic energy dominated
regime. This is of relevance to parton scattering in the
LHC if large extra dimension scenarios describe these
interactions in the TeV regime, for then high energy partons
could form black holes, and the scattering process would
be well approximated by the collision of two black holes.
We presented a sketch of how the geodesic analogue
could be used to estimate the black hole formation 
cross section, and energy radiated to gravitational
waves as a function of the impact parameter. At threshold
it is conceivable that essentially {\em all} the kinetic
energy is radiated as gravitational waves. Away
from threshold significant amounts of energy could
still be radiated even if a black hole does not form.

Clearly, much of the above is pure speculation, though 
we believe is sufficiently interesting and relevant
that it will be a fruitful endeavor to try to further establish
(or disprove) the correspondence between geodesic 
scattering and the full problem. Simulating the ultra relativistic
collision of black holes will computationally be a very
challenging problem, and probably not possible
to do beyond the 4-dimensional case without peta-flop scale
computing. Though if the correspondence could be establish
more strongly in the 4-dimensional case (which could be tackled
with tera-flop resources, at least for moderately relativistic
energies), then comparable mass head-on collision simulations
in higher dimensions together with studies of geodesic scattering
and estimates of energy and angular momentum emission in gravitational
waves, similar to that performed in~\cite{berti_et_al} for example,
could be used to provide useful information for the higher dimensional
case.

\noindent{\bf{\em Acknowledgments:}}
FP would like to thank Don Page, Eric Poisson,
Hirotaka Yoshino and Douglas Gingrich for insightful discussions.
FP gratefully acknowledges research support from the CIAR,
NSERC and Alberta Ingenuity.
The simulations described here were performed on 
the University of British Columbia's {\bf vnp4}
cluster (supported by CFI and BCKDF), {\bf WestGrid} machines
(supported by CFI, ASRI and BCKDF), the Dell {\bf Lonestar} cluster
at the University of Texas in Austin, {\bf Sharcnet} facilities 
(principle support by CFI, OIT and ORDCF) and the {\bf McKenzie} cluster
at CITA (supported by the CFI and OIT).

\appendix

\section{Geodesic Integration}\label{sec_geod_details}

In this appendix we describe the manner in which we integrate
geodesics in the Kerr geometry. The original purpose
of this geodesic integration method was for straight-forward
incorporation into the GH evolution code, to study geodesic
propagation in binary merger spacetimes. Thus the method
does not take advantage of any of the symmetries of the underlying
spacetime, nor uses advanced high-order ordinary differential
equation (ODE) integrators. Nevertheless given the speed of contemporary
desktop PC's and that geodesic integration is a one dimensional evolution 
there is no problem in obtaining sufficient accuracy for the purposes
of the studies presented in the main sections of the paper.

Consider a spacetime with metric $g_{\alpha\beta}$
\be
ds^2=g_{\alpha\beta}dx^\alpha dx^\beta,
\ee
and a geodesic of the spacetime described via the
parametric representation of its curve 
${x}^\alpha={x}^\alpha(\lambda)$,
where $\lambda$ is the affine parameter along the curve. The geodesic's
tangent vector is
\be
u^\alpha=\frac{d x^\alpha(\lambda)}{d\lambda} \equiv x^{\prime\alpha},\label{udef}
\ee
defining prime ($\prime$) to denote differentiation with respect to $\lambda$.
For ${x}^\alpha$ to represent a geodesic, $u^\alpha$ must satisfy
the geodesic equation:
\be
u^{\prime\alpha} + \Gamma_{\gamma\delta}^\alpha 
u^{\gamma} u^{\delta}=0,\label{gc}
\ee
where $\Gamma_{\gamma\delta}^\alpha$ are the Christoffel symbols (\ref{christoff}).
In terms of the
coordinate position along the curve ${x}^\alpha(\lambda)$,
(\ref{gc}) can be written as:
\be
x^{\prime\prime\alpha} + \Gamma_{\gamma\delta}^\alpha 
x^{\prime\gamma} x^{\prime\delta}=0.\label{gc2}
\ee
The causal character of the geodesic is given by the normalization
of its tangent vector, as follows:
\bea
u^\alpha u^\beta g_{\alpha\beta} &=& -1 \ \ \ {\rm timelike}\label{tl}\\
u^\alpha u^\beta g_{\alpha\beta} &=&\ 0 \ \ \ {\rm null}\label{null}\\
u^\alpha u^\beta g_{\alpha\beta} &=&\ 1 \ \ \ {\rm spacelike}
\eea
We are most interested in timelike and null geodesics at the moment.
For timelike geodesics, the above normalization is equivalent
to demanding that $\lambda$ measures the proper time of an
observer moving along the geodesic. $\lambda$ does not have such
a straight-forward interpretation for a null geodesic; in fact,
we can re-parameterize any null geodesic via a linear transformation
of the form $s=a\lambda+b$, where $a$ and $b$ are constants. 

The geodesic equations (\ref{gc2}) are a set of four ODEs
for the coordinate position of the
corresponding ``particle'' as a function of affine time $\lambda$.
There are at least a couple of ways to proceed to solve these
equations. One is to integrate (\ref{gc2}) directly as a set
of second order ODEs, the other is to reduce it to a system
of first order ODEs. We will take the former approach as
it will fit ``naturally'' within the generalized harmonic evolution
code metrics that we want to explore the geodesic structure of.
Another practical consideration in the code is that
we need to integrate with respect to coordinate time $t$ and {\em not}
affine time $\lambda$. Define $x^0=t$, and let $x^k$, $k\in1,2,3$ be
the three spatial coordinates, i.e. $(x^1,x^2,x^3)=(x,y,z)$.
We thus want to solve for $x^k(t)$
for each geodesic. With the overdot ($\dot{\ }$) denoting differentiation with
respect to $t$, using the chain rule we get:
\bea
x^{\prime k} = \dot{x}^k t^\prime \label{c1} \\\ 
x^{\prime\prime k} = \ddot{x}^k t^{\prime 2} + \dot{x}^k t^{\prime\prime}, \label{c2}
\eea
Substituting (\ref{c1},\ref{c2}) into (\ref{gc2}) gives
\be
\ddot{x}^\alpha t^{\prime 2} + \dot{x}^\alpha t^{\prime\prime} + 
\Gamma_{\gamma\delta}^a \dot{x}^\gamma \dot{x}^\delta t^{\prime 2} =0.\label{gc3}
\ee
Equation (\ref{gc2}) for $t$ reads:
\be
t^{\prime\prime} + \Gamma_{\gamma\delta}^0 x^{\prime\gamma} x^{\prime\delta}=0,
\ee
and using this, with (\ref{c1}) again,
the geodesic equation (\ref{gc3}) for the spatial coordinates $x^k$ can be written as:
\be
\ddot{x}^k + \left( \Gamma_{\gamma\delta}^k - \dot{x}^k \Gamma_{\gamma\delta}^0\right) 
\dot{x}^\gamma \dot{x}^\delta = 0. \label{geq}
\ee
This is the set of equations we will solve for timelike and null geodesics
in a general spacetime $g_{\alpha\beta}$.

\subsection{Initial Conditions}\label{sec_ic}
Since we are solving three, second order in time ODEs for each geodesic,
we need {\em six} initial conditions per curve: the initial position
$x^k(t=0)$ and coordinate velocity $\dot{x}^k(t=0)$ of each particle. Note
that we cannot choose arbitrary velocities, as our choices must
be consistent with the normalization conditions (\ref{tl},\ref{null}). There
are several conceivable ways to ensure this; we will take the following route.

We begin by choosing
some arbitrary direction $k^\gamma = (0,k^x,k^y,k^z)$ for the curve to point
in. We need the {\em unit} spatial vector in this direction; call it $\hat{k}^\gamma$:
\be\label{knorm}
\hat{k}^\gamma \hat{k}^\delta g_{\gamma\delta} = 1
\ee
and $\hat{k}^\gamma$ can be found by defining 
\be
\hat{k}^\gamma = N k^\gamma,
\ee
plugging this into the normalization condition (\ref{knorm}), and solving for $N$.
The tangent vector $u^\alpha$ corresponding to a photon moving in the direction 
$k^\gamma$ is then:
\be
u^\alpha = n^\alpha + \hat{k}^\alpha \ \ \ (null\ case) \label{null_ic}
\ee
For a timelike curve, in addition to the direction $\hat{k}^\alpha$
we can choose a velocity $v$ ($<1$). The corresponding tangent vector in this case
is
\be
u^\alpha = \Gamma \left (n^\alpha + v \hat{k}^\alpha\right) \ \ \ (timelike\ case),\label{tl_ic}
\ee
where $\Gamma$ is the Lorentz gamma factor (here relative to an observer
sitting at constant coordinate location and thus moving in the $n^\alpha$ direction):
\be
\Gamma=\frac{1}{\sqrt{1-v^2}}.
\ee
Now that we know $u^\alpha$, using (\ref{udef}) and (\ref{c1}) we can find the 
initial coordinate velocities of our geodesic curves. 

\subsection{Numerical Technique}

For compatibility with the metric evolution code we use a three time level
scheme in a Cartesian coordinate system. The discrete version of the curve
$(x(t),y(t),z(t))$ is represented by ${x^i,y^i,z^i}$, where
the time $t^i=i\Delta t$, with $\Delta t$ the discretization scale.
Time derivatives are computed using standard second order 
accurate stencils:
\bea
\dot{f}(t)|_{t=t^i} = \frac{f^{i+1} - f^{i-1}}{2 \Delta t} + O(\Delta t^2)\nonumber\\ 
\ddot{f}(t)|_{t=t^i} = \frac{f^{i+1} - 2 f^i + f^{i-1}}{\Delta t^2} + O(\Delta t^2).\label{tdiff}
\eea

Thus, when evolving from $t=t^i$ to $t=t^{i+1}$, we require the locations
of the geodesics at $t=t^i$ and $t=t^{i-1}$. In the geodesic equation (\ref{geq})
the Christoffel symbols are supplied as known functions of the spacetime.

When the time differences (\ref{tdiff}) are substituted into the geodesic
equation (\ref{geq}) for the time derivatives of the three coordinate positions
of a geodesic, we end up with an algebraic system of three equations for three
unknowns: $(x^{i+1},y^{i+1},z^{i+1})$. These equations are non-linear, and
so an efficient method to solve for them is via Newton iteration.
We write the system of equations as 
\be
\Screll_j(q^k)=0,\label{eqns}
\ee
where $j=1,2,3$ labels one of the equations, and we now use the notation
$q^k$, $k=1,2,3$ to label one of the unknowns (i.e. $q^1=x^{i+1},q^2=y^{i+1},q^3=z^{i+1}$).
Newton iteration then proceeds by writing the 
unknowns as a guess $\hat{q}_1^k$ plus a correction $\delta q^k$,
\be
q^k = \hat{q}_1^k + \delta q^k,
\ee
linearizing about the guess, and solving for the corrections
to first order:
\bea
\Screll_j(q^k) &=& \Screll_j(\hat{q}_1^k + \delta q^k) \nonumber\\
               &\approx& \Screll_j|_{q^k=\hat{q}_1^k} + \frac{\partial\Screll_j}{\partial q^l }|_{q^k=\hat{q}_1^k}\cdot \delta q^l = 0.
\label{lin_resid}
\eea
Written in terms of the residual $R_j\equiv \Screll_j(\hat{q}_1^k)$
and the Jacobian matrix $J_{jl}\equiv \frac{\partial\Screll_j}{\partial q^l }$
the resultant linear system that is solved for the correction $\delta_q^l$ is:
\be
J_{jl} \cdot \delta q^l = -R_j.
\ee
After solving for the correction, the above steps are repeated with the
corrected solution serving as the new guess, and the iteration
proceeds until the norm of the residual is below some specified tolerance.

\subsubsection{Second order accurate initial conditions}\label{sec_2nd_ic}
To begin evolving the geodesics at $t=0$ with a three time level scheme,
we need the initial positions $x^0,y^0,z^0$ at $t=0$ {\em and} the
positions $x^{-1},y^{-1},z^{-1}$ at $t=-\Delta t$. One can use taylor expansions,
the freely specifiable initial conditions discussed in Sec.~\ref{sec_ic},
and the geodesic equations (\ref{geq}) to initialize the past time values
to the necessary level of accuracy. Use the subscript $0$ to refer to
the analytic initial condition for a given variable; for example,
for $x$:
\bea
x_0 \equiv x(t)|_{t=0},\\
\dot{x}_0 \equiv \dot{x}(t)|_{t=0},\\
\ddot{x}_0 \equiv \ddot{x}(t)|_{t=0}.
\eea
Then, to second order
\bea
x^{-1} &=& x_0 - \dot{x}_0 \Delta t + \ddot{x}_0 \frac{\Delta t^2}{2}.
\eea
The initial position $x^k_0$ is freely specifiable, the initial
velocity $\dot{x}^k_0$ is calculated as described in (\ref{sec_ic}),
and the geodesic equations (\ref{geq}) are used to solve for  $\ddot{x}^k_0$:
\be
\ddot{x}^k_0 = - \left(\Gamma_{\gamma\delta}^k - \dot{x}^k_0 \Gamma_{\gamma\delta}^t\right) \dot{x}_0^{\gamma} \dot{x}_0^{\delta}.  \\
\ee
Note the summation over all initial velocities in the above, including 
over $\dot{t} = 1$.

\end{document}